\documentclass[Physsubmission, Phys]{SciPost}

\usepackage[english]{babel}

\hypersetup{
    colorlinks,
    linkcolor={red!50!black},
    citecolor={blue!50!black},
    urlcolor={blue!80!black}
}

\usepackage[bitstream-charter]{mathdesign}
\DeclareSymbolFont{usualmathcal}{OMS}{cmsy}{m}{n}
\DeclareSymbolFontAlphabet{\mathcal}{usualmathcal}

\begin{document}

\begin{center}{\Large \textbf{
Recent results on hyperon pair production and nucleon time-like form factors at BESIII\\
}}\end{center}

\begin{center}
Alessio~Mangoni\textsuperscript{1}\\
{\it {\footnotesize on behalf of the BESIII Collaboration}}
\end{center}

\begin{center}
{\bf 1} Istituto Nazionale di Fisica Nucleare, Sezione di Perugia, Italy
\\
* alessio.mangoni@pg.infn.it
\end{center}

\begin{center}
\today
\end{center}


\definecolor{palegray}{gray}{0.95}
\begin{center}
\colorbox{palegray}{
  \begin{tabular}{rr}
  \begin{minipage}{0.1\textwidth}
    \includegraphics[width=30mm]{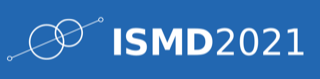}
  \end{minipage}
  &
  \begin{minipage}{0.75\textwidth}
    \begin{center}
    {\it 50th International Symposium on Multiparticle Dynamics}\\ {\it (ISMD2021)}\\
    {\it 12-16 July 2021} \\
    \doi{10.21468/SciPostPhysProc.?}\\
    \end{center}
  \end{minipage}
\end{tabular}
}
\end{center}

\section*{Abstract}
{\bf
The BESIII experiment, based on the BEPCII collider at IHEP in Beijing, China, is capable of measuring nucleon form factors with unprecedented precision. We report recent results achieved by the BESIII collaboration concerning the nucleon time-like form factors and the hyperon pair production. Specifically, we outline the results obtained by the study of the process $e^+e^- \to p \overline p$, using both the scan-energy and initial-state radiation techniques, those of the process $e^+e^- \to n \overline n$, and the study of the $\Sigma$ polarization regarding the process $e^+e^- \to J/\psi \to \Sigma^+\overline \Sigma{}^-$.
}

\vspace{10pt}
\noindent\rule{\textwidth}{1pt}
\tableofcontents\thispagestyle{fancy}
\noindent\rule{\textwidth}{1pt}
\vspace{10pt}

\section{The BESIII experiment}
The BESIII experiment is based on the BESIII (Beijing Spectrometer III) detector and BEPCII (Beijing Electron-Positron Collider II) accelerator located at IHEP in Beijing, China~\cite{BESIII:2020nme}. \\
The physics program covers the following topics: tests of electroweak interactions; studies of light hadron spectroscopy and decay properties; studies of the production and decay properties of the main charmonia; studies of charm and $\tau$-physics; search for glueballs, quark-hybrids, multi-quark states and other exotic states; precision measurements of QCD and CKM parameters; and searches for new physics.
The first version of BEPC dates back to 1984 and the first operation started in 1989. It was subsequently upgraded to BEPCII in 2008. The BEPCII collider is a double-ring collider with a design luminosity of $1 \times 10^{33} \ \rm cm^{-2} s^{-1}$, reached in 2016~\cite{BESIII:2020nme}, and designed originally to collide $e^+e^-$ beams in the energy range $($2-4.6$) \ \rm GeV$. The beam energy is tunable and, in February 2021, the center-of-mass energy reached the record of $4.946 \ \rm GeV$. Further details can be found in Refs.~\cite{BESIII:2009fln,Yuan:2019zfo,BESIII:2020nme}. The details of the BESIII detector can be found in Ref.~\cite{BESIII:2009fln}.\\
The BESIII collaboration has about 500 members from 78 institutions and 16 countries. The recent remarkable results include the accumulation of 10 billion $J/\psi$ events and the observation of polarization of baryons in $J/\psi$ decays, based on 1.3 billion $J/\psi$ candidates~\cite{BESIII:2018cnd}.

\section{Nucleon electromagnetic form factors}
\begin{figure}[h]
\centering
\includegraphics[width=0.9\textwidth]{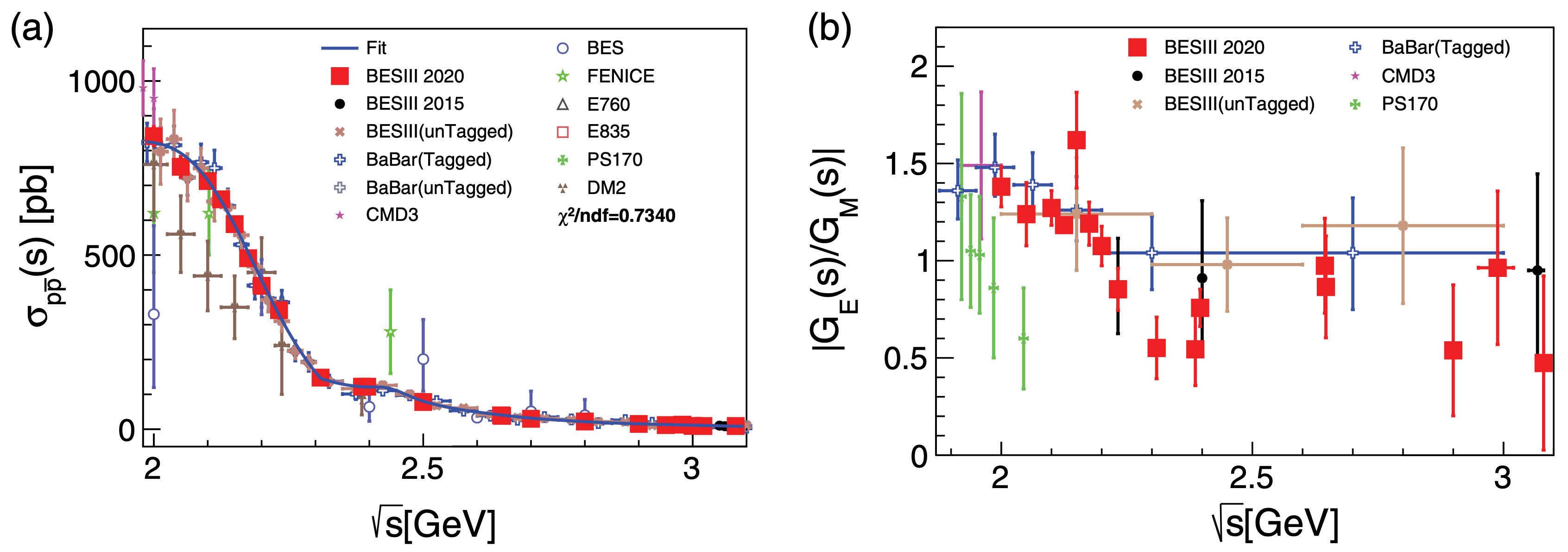}
\caption{BESIII results for the cross section of $e^+e^- \to p \overline p$  (red solid squares), panel (a), and the ratio $|G_E/G_M|$ of the proton, panel (b)~\cite{BESIII:2019hdp}.}
\label{fig.bppmr}
\end{figure}

\begin{figure}[h]
\centering
\includegraphics[width=0.95\textwidth]{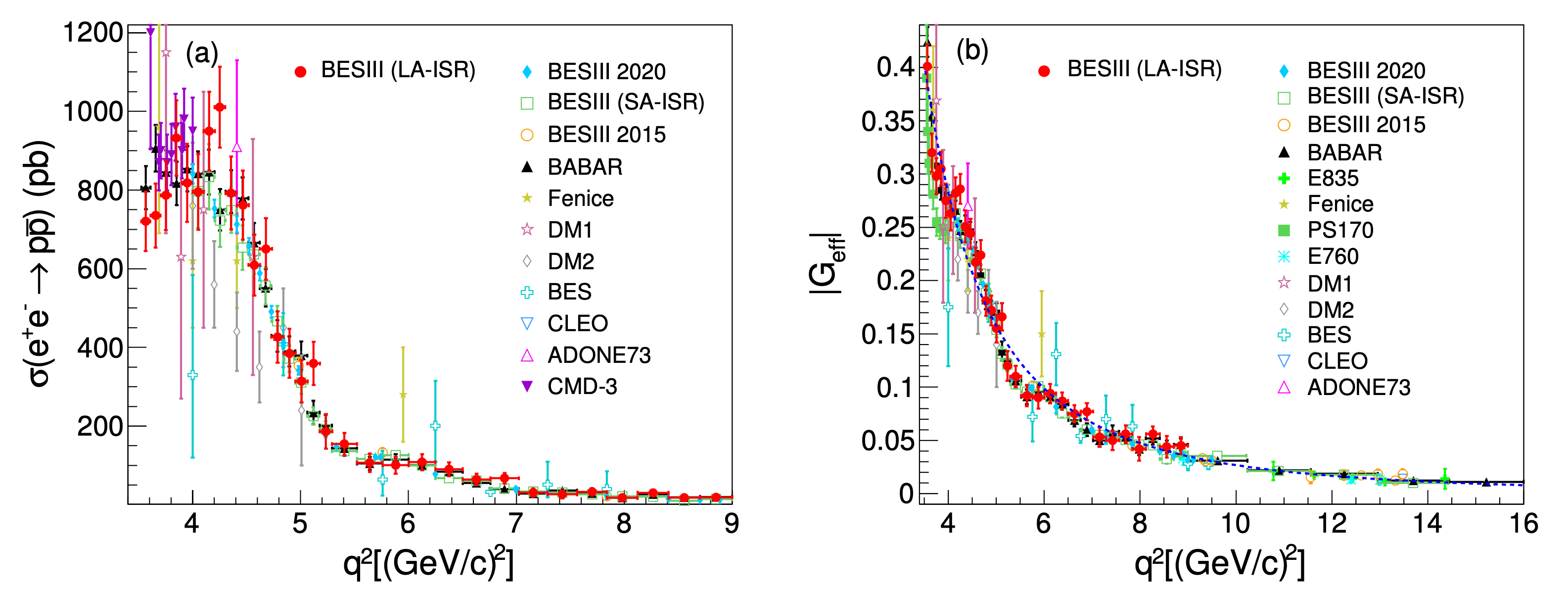}
\caption{BESIII results for the cross section of the process $e^+e^- \to p \overline p$ and the effective FF of the proton~\cite{BESIII:2021rqk}.}
\label{fig.bppcsmr}
\end{figure}

\begin{figure}[h]
\centering
\includegraphics[width=0.95\textwidth]{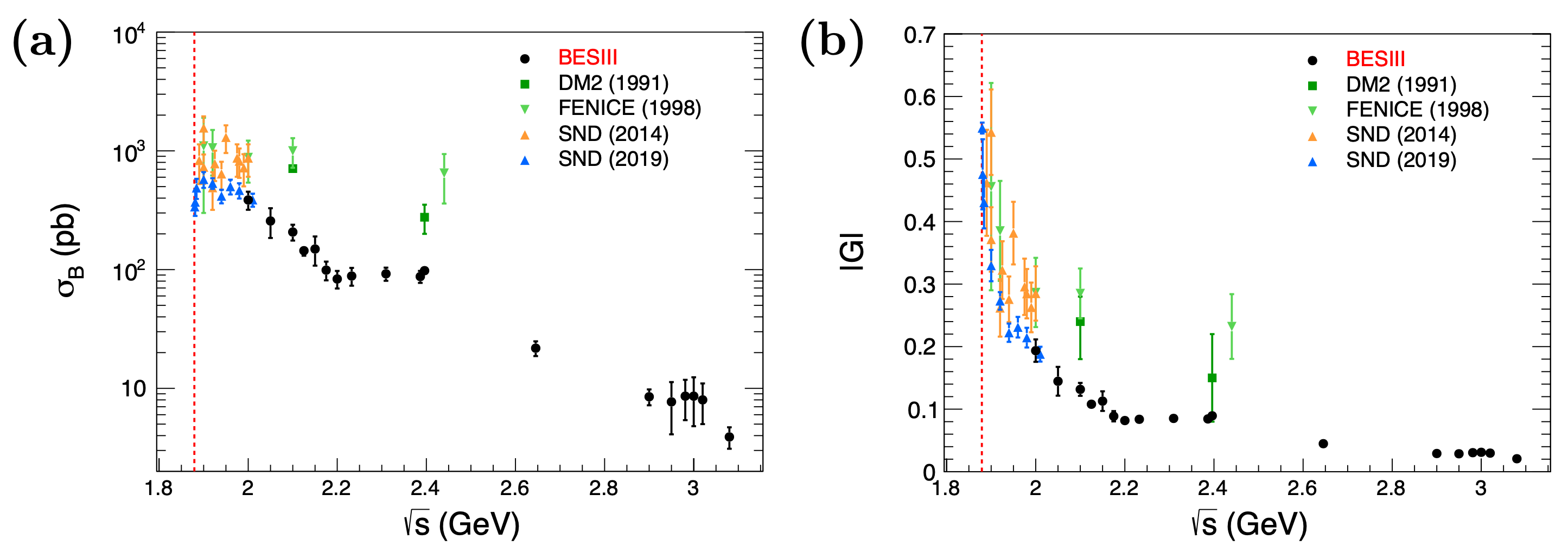}
\caption{Cross section, panel (a), and effective form factor, panel (b), for the process $e^+e^- \to n \overline n$~\cite{BESIII:2021dfy}.}
\label{fig.bnnmr}
\end{figure}

The nucleon electromagnetic (EM) form factors (FFs) are Lorentz scalar functions of $q^2$ (squared four-momentum transfer of the photon) and give informations about the internal structure of nucleons.
\begin{figure}[h]
\centering
\includegraphics[width=0.6\textwidth]{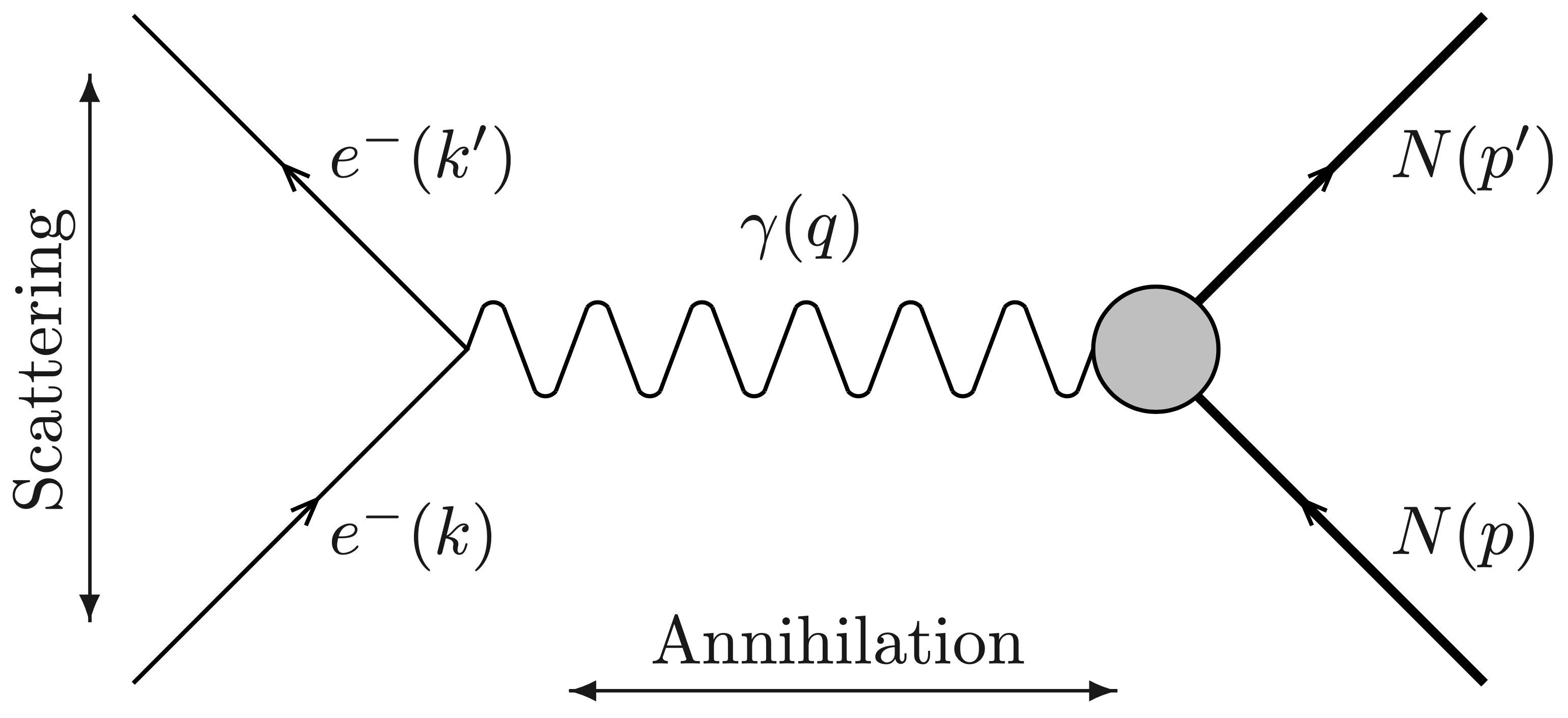}
\caption{Feynman diagram for scattering and annihilation.}
\label{fig.feyn}
\end{figure}\\
Space-like FFs are related to the elastic scattering process of the kind $e^- N \to e^- N$, while time-like FFs are related to the annihilation processes, such as $e^+e^- \leftrightarrow N \overline N$, the Feynman diagram of which is shown in Figure~\ref{fig.feyn}. We can introduce two independent EM FFs, the Sachs electric and magnetic FFs, defined as
$$
G_E(q^2)=F_1(q^2)+ {q^2 \over 4M_N} F_2(q^2) \,, \ \ \ \ \ \ \ \ \ \ G_M(q^2)=F_1(q^2)+  F_2(q^2) \,,
$$
where $M_N$ is the mass of nucleon and $F_1(q^2)$ and $F_2(q^2)$ are the the Dirac and Pauli FFs, respectively.\\
Concerning the recent experimental results, in 2020 the BESIII collaboration has performed accurate measurements of the cross section for the process $e^+e^- \to p \overline p$ with center of mass (CM) energy from 2.0 GeV to 3.08 GeV, for a total of 22 points. The cross section is measured with the energy-scan technique, using 688.5 pb$^{-1}$ of integrated luminosity~\cite{BESIII:2019hdp}. The modulus of the ratio $G_E/G_M$ has been obtained with high accuracy. The cross section and the modulus of $G_E/G_M$ results are reported in Figure~\ref{fig.bppmr}, in panel (a) and (b), respectively.
\\
The value of $|G_E|$ and $|G_M|$ of the proton have been separated for the first time
and an oscillating behavior is observed in the effective FF as a function of the relative momentum $p$
between proton and anti-proton.
This behavior has been observed also by further BESIII and BABAR analyses, and the most probable explanations are the presence of resonant structures~\cite{Lorenz:2015pba} or interference effects in the final state~\cite{Tomasi-Gustafsson:2020vae}.\\
The process $e^+e^- \to p \overline p$ has been studied by the BESIII collaboration, also with initial-state radiation (ISR) technique, using 7.5 fb$^{-1}$ of integrated luminosity at 7 energy points from 3.773 to 4.600 GeV, obtaining measurements of the cross section from the production threshold to 3.0 GeV~\cite{BESIII:2021rqk}. The obtained results for the cross section and effective FF are reported in panels (a) and (b) of Figure~\ref{fig.bppcsmr}, respectively. 
\\
Very recently, the BESIII collaboration have analyzed also the $e^+e^- \to n \overline n$ process with the measurement of the cross section from 2.0 GeV to 3.08 GeV, for 18 energy points in total, corresponding to an integrated luminosity of 647.9 pb$^{-1}$~\cite{BESIII:2021dfy}. The best precision reached for the cross section is about of 8.1\% at the energy of 2.396 GeV. The obtained results for cross section and effective FF are shown in Figure~\ref{fig.bnnmr}, panel (a) and (b), respectively.
Oscillations in the modulus of the effective FF, after the subtraction of a dipole function, is observed, similar to the case of a proton, as shown in Figure~\ref{fig.bnnmr}.
\begin{figure}[h]
\centering
\includegraphics[width=0.6\textwidth]{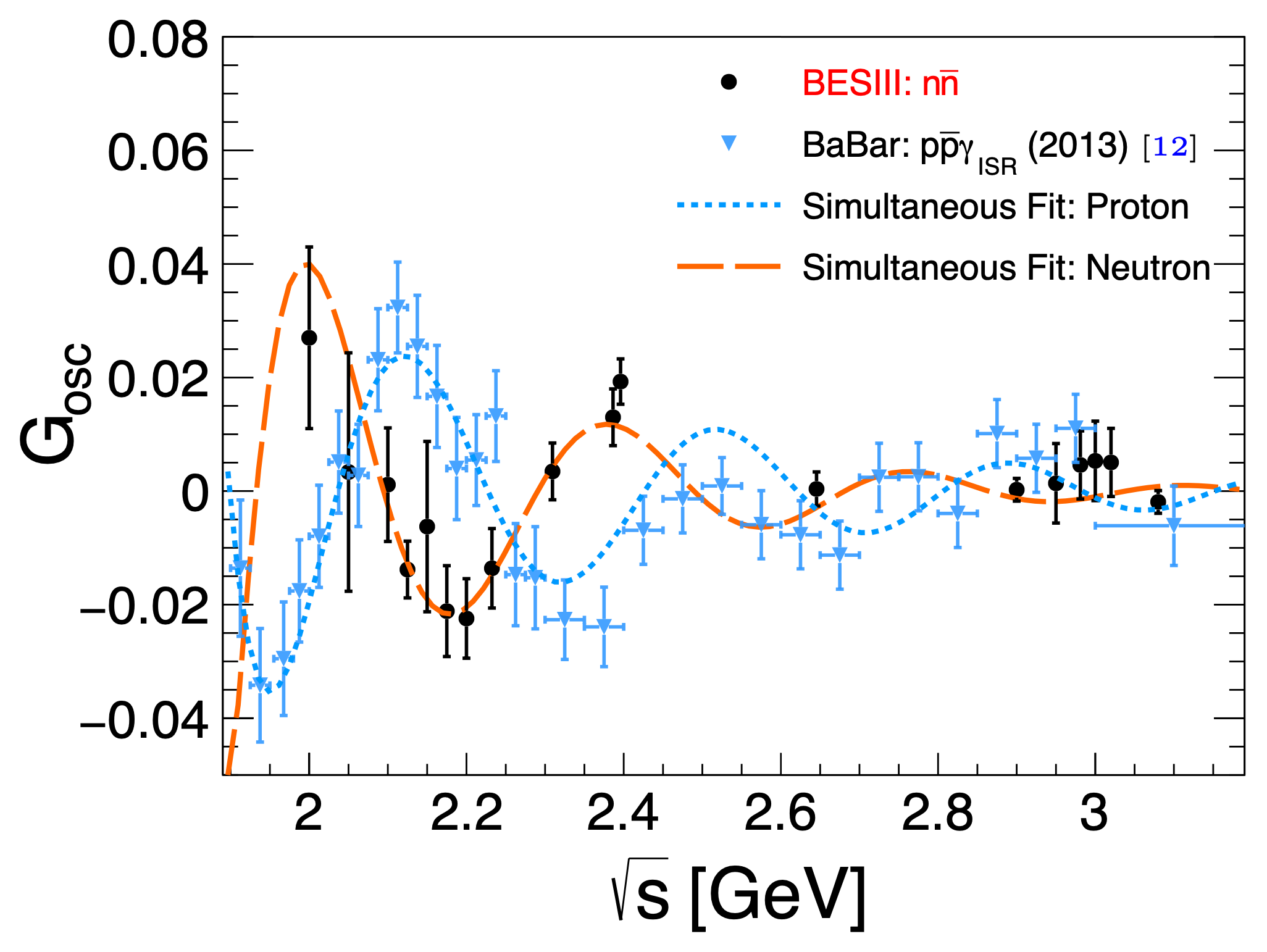}
\caption{Oscillating behavior of neutron effective FF~\cite{BESIII:2021dfy}.}
\label{fig.bnnmfosc}
\end{figure}\\
The oscillating function $G_{\rm osc}$ has the form
$$
G_{\rm osc}(q^2) = |G|-G_D(q^2) \,, \ \ \ \ \ \ G_D(q^2) = {\mathcal A_n \over \left(1-{q^2 \over 0.71 \ { \rm GeV^2}}\right)^2}\,, \ \ \ \ \ \ A_n = 4.87 \pm 0.09\,,
$$
where $|G|$ is the modulus of the effective FF and $G_D$ is the dipole function.

\section{Hyperon pair production}
Hyperons are ideal probes for studying the strong interaction in the transition region, where the regime is not completely perturbative. \\
Considering the process $e^+e^- \to J/\psi \to \Sigma^+(p \pi^0) \overline \Sigma{}^- (\overline p \pi^0)$, the polarization of hyperons can be determined by analyzing the two-body weak decays $\Sigma^+ \to p \pi^0 $ and $\overline \Sigma{}^- \to \overline p \pi^0$. In particular, the $e^+e^- \to \psi \to \Sigma^+ \overline \Sigma{}^-$ production process is described by the psionic electric and magnetic FFs $G_E^\psi$ and $G_M^\psi$ related to the parameters $\alpha_\psi$ (angular decay asymmetry parameter) and $\Delta \Phi$ (relative phase between FFs).\\
The polarization parameters $\alpha_0$, $\overline \alpha_0$ and the relative phase between FFs, $\Delta \Phi$, have been determined by the BESIII collaboration. In particular, the value of $\overline \alpha_0 = 0.990 \pm 0.037 \,(\rm stat) \pm 0.011\, (\rm syst)$ has been measured for the first time. The results are reported in Table~\ref{tab.1} and are based on $1310.6 \times 10^{6} J/\psi$ and $448.1 \times 10^{6} \psi(2S)$ events collected by the BESIII collaboration~\cite{BESIII:2020fqg}.
\begin{table}[]
\centering
\begin{tabular}{|c|c|}
\hline
\textbf{Parameter} & \textbf{Measured value}  \\ \hline
$\alpha_{J/\psi}$ & $-0.508\pm 0.006\pm0.004$ \\ \hline
$\Delta \Phi_{J/\psi}$ & $-0.270\pm 0.012\pm0.009$ \\ \hline
$\alpha_{\psi(2S)}$ & $0.682\pm 0.03\pm0.011$   \\ \hline
$\Delta \Phi_{\psi(2S)}$ & $0.379\pm 0.07\pm0.014$   \\ \hline
$\alpha_0$ 	& $-0.998\pm0.037\pm0.009$ \\ \hline
$\overline \alpha_0$ & $0.990\pm 0.037\pm 0.011$  \\ \hline
\end{tabular}
\caption{\label{tab.1}BESIII results~\cite{BESIII:2020fqg}.}
\end{table}
Concerning the study of the processes $e^+e^- \to \Sigma^+ \overline \Sigma{}^-$ and $e^+e^- \to \Sigma^- \overline \Sigma{}^+$, BESIII have performed measurements of the cross section for CM energies ranging from 2.3864 GeV to 3.0200 GeV. 
The results represent the first measurements of the cross section in the off-resonance region~\cite{BESIII:2020uqk}.
The study of the ratio of the $\Sigma$ hyperon FFs could provide guidance for the nucleons, since their valence quark composition is similar, with a replacement of the $s$ quark in the $\Sigma$ baryons, with the $d$ and $u$ quarks for the proton and neutron, respectively.\\
The obtained value of the modulus of the ratio $G_E/G_M$ is
$$
\left|{G_E \over G_M}\right| = 1.83 \pm 0.26 
$$
for the $\Sigma^+$ hyperon, at the energy $\sqrt s = 2.3960 \ \rm GeV$, is significantly higher than 1. Moreover, the cross section values near threshold disagree with the point-like expectations, as seen for the proton, see Refs.~\cite{Baldini:2007qg,BaldiniFerroli:2010ruh,BaBar:2013ves,CMD-3:2015fvi}.\\
The obtained value of the ratio
$$
{\sigma^{\rm Born}(e^+e^- \to \Sigma^+ \overline \Sigma{}^-) \over \sigma^{\rm Born}(e^+e^- \to \Sigma^- \overline \Sigma{}^+)} = 9.7 \pm 1.3
$$
is consistent with the prediction of Ref.~\cite{BaldiniFerroli:2019abd}, but inconsistent with other predictions from various models~\cite{Chernyak:1983ej,Anselmino:1992vg,Jaffe:2003sg,Ramalho:2019koj}.

\section{Conclusion}
The BESIII experiment has the ability to precisely measure the nucleon FFs. The $e^+e^- \to p \overline p$ process has been studied to obtain the differential cross section, the EM FFs, $|G_E|$ and $|G_M|$, and their ratio $|G_E/G_M|$. The moduli of electric and magnetic proton FFs has been measured for the first time for the proton. Concerning the neutron, the $e^+e^- \to n \overline n$ process has been studied to obtain the cross section and the ratio of EM FFs, $|G_E/G_M|$, with high precision. An oscillating behavior is observed in the effective FF of both the proton and neutron. Moreover, the processes of hyperon pairs production can be used to study of hyperon-pair production. BESIII has measured for the first time all the $\Sigma$ polarization parameters, by studying the process $e^+e^- \to J/\psi \to \Sigma^+ \overline \Sigma{}^-$ and the subsequent decays $\Sigma^+ \to p \pi^0$ and $\overline \Sigma{}^- \to \overline p \pi^0$.

\section*{Acknowledgements}
This work was supported in part by the STRONG-2020 project of the European Union’s Horizon 2020 research and innovation programme under Grant agreement no. 824093.

\nolinenumbers


\begin{thebibliography}{}

\bibitem{BESIII:2020nme}
M.~Ablikim \textit{et al.} [BESIII],
Chin. Phys. C \textbf{44}, no.4, 040001 (2020)

\bibitem{BESIII:2009fln}
M.~Ablikim \textit{et al.} [BESIII],
Nucl. Instrum. Meth. A \textbf{614}, 345-399 (2010)

\bibitem{Yuan:2019zfo}
C.~Z.~Yuan and S.~L.~Olsen,
Nature Rev. Phys. \textbf{1}, no.8, 480-494 (2019)

\bibitem{BESIII:2018cnd}
M.~Ablikim \textit{et al.} [BESIII],
Nature Phys. \textbf{15}, 631-634 (2019)

\bibitem{BESIII:2019hdp}
M.~Ablikim \textit{et al.} [BESIII],
Phys. Rev. Lett. \textbf{124}, no.4, 042001 (2020)

\bibitem{Lorenz:2015pba}
I.~T.~Lorenz, H.~W.~Hammer and U.~G.~Mei\ss{}ner,
Phys. Rev. D \textbf{92}, no.3, 034018 (2015)

\bibitem{Tomasi-Gustafsson:2020vae}
E.~Tomasi-Gustafsson, A.~Bianconi and S.~Pacetti,
Phys. Rev. C \textbf{103}, no.3, 035203 (2021)


\bibitem{BESIII:2021rqk}
M.~Ablikim \textit{et al.} [BESIII],
Phys. Lett. B \textbf{817}, 136328 (2021)

\bibitem{BESIII:2021dfy}
M.~Ablikim \textit{et al.} [BESIII],
[arXiv:2103.12486 [hep-ex]].


\bibitem{BESIII:2020fqg}
M.~Ablikim \textit{et al.} [BESIII],
Phys. Rev. Lett. \textbf{125}, no.5, 052004 (2020)

\bibitem{BESIII:2020uqk}
M.~Ablikim \textit{et al.} [BESIII],
Phys. Lett. B \textbf{814}, 136110 (2021)

\bibitem{Baldini:2007qg}
R.~Baldini, S.~Pacetti, A.~Zallo and A.~Zichichi,
Eur. Phys. J. A \textbf{39}, 315-321 (2009)

\bibitem{BaldiniFerroli:2010ruh}
R.~Baldini Ferroli, S.~Pacetti and A.~Zallo,
Eur. Phys. J. A \textbf{48}, 33 (2012)

\bibitem{BaBar:2013ves}
J.~P.~Lees \textit{et al.} [BaBar],
Phys. Rev. D \textbf{87}, no.9, 092005 (2013)

\bibitem{CMD-3:2015fvi}
R.~R.~Akhmetshin \textit{et al.} [CMD-3],
Phys. Lett. B \textbf{759}, 634-640 (2016)

\bibitem{BaldiniFerroli:2019abd}
R.~Baldini Ferroli, A.~Mangoni, S.~Pacetti and K.~Zhu,
Phys. Lett. B \textbf{799}, 135041 (2019)

\bibitem{Chernyak:1983ej}
V.~L.~Chernyak and A.~R.~Zhitnitsky,
Phys. Rept. \textbf{112}, 173 (1984)

\bibitem{Anselmino:1992vg}
M.~Anselmino, E.~Predazzi, S.~Ekelin, S.~Fredriksson and D.~B.~Lichtenberg,
Rev. Mod. Phys. \textbf{65}, 1199-1234 (1993)

\bibitem{Jaffe:2003sg}
R.~L.~Jaffe and F.~Wilczek,
Phys. Rev. Lett. \textbf{91}, 232003 (2003)

\bibitem{Ramalho:2019koj}
G.~Ramalho, M.~T.~Pe\~na and K.~Tsushima,
Phys. Rev. D \textbf{101}, no.1, 014014 (2020)




	
\end{thebibliography}
\end{document}